\documentclass[a4paper,10pt]{IEEEtran}
\renewcommand{\baselinestretch}{0.99}
\usepackage{pgfplots}
\usetikzlibrary{shapes.multipart,intersections}
\usepackage{cite}
\usepackage{amsmath,amssymb,amsfonts,steinmetz}
\usepackage{algorithmic}
\usepackage{graphicx}
\usepackage{textcomp}
\usepackage{xcolor,flushend}
\def\BibTeX{{\rm B\kern-.05em{\sc i\kern-.025em b}\kern-.08em
    T\kern-.1667em\lower.7ex\hbox{E}\kern-.125emX}}

\usepackage{tikz}
\usetikzlibrary{calc}
\makeatletter
\newcommand{\gettikzxy}[3]{%
  \tikz@scan@one@point\pgfutil@firstofone#1\relax
  \edef#2{\the\pgf@x}%
  \edef#3{\the\pgf@y}%
}
\makeatother

\usepackage[T1]{fontenc}
\usepackage[latin9]{inputenc}
\usepackage{bm}
\usepackage{amsmath}
\usepackage{amssymb}
\usepackage{babel}
\usepackage{comment}
\usepackage{subcaption}
\usepackage{graphicx}
\usepackage{algorithm,algorithmic}
\usepackage{xcolor}
\usepackage{gensymb}
\usepackage{cite}
\usepackage{enumitem}
\usepackage{url}

\renewcommand{\a}{\mathbf{a}}

\newcommand{\f}{\mathbf{f}}
\newcommand{\g}{\mathbf{g}}
\newcommand{\h}{\mathbf{h}}

\newcommand{\n}{\mathbf{n}}

\newcommand{\p}{\mathbf{p}}

\newcommand{\s}{\mathbf{s}}

\renewcommand{\u}{\mathbf{u}}
\renewcommand{\v}{\mathbf{v}}

\newcommand{\x}{\mathbf{x}}
\newcommand{\y}{\mathbf{y}}
\newcommand{\z}{\mathbf{z}}

\newcommand{\A}{\mathbf{A}}
\newcommand{\B}{\mathbf{B}}
\newcommand{\C}{\mathbf{C}}
\newcommand{\D}{\mathbf{D}}

\newcommand{\F}{\mathbf{F}}

\renewcommand{\H}{\mathbf{H}}
\newcommand{\I}{\mathbf{I}}
\newcommand{\J}{\mathbf{J}}
\newcommand{\K}{\mathbf{K}}

\newcommand{\N}{\mathbf{N}}

\renewcommand{\P}{\mathbf{P}}
\newcommand{\Q}{\mathbf{Q}}

\renewcommand{\S}{\mathbf{S}}
\newcommand{\T}{\mathbf{T}}

\newcommand{\W}{\mathbf{W}}

\newcommand{\Y}{\mathbf{Y}}

\newcommand{\Compl}{\mbox{$\mathbb{C}$}}

\renewcommand{\Re}{\mathrm{Re}}

\usepackage{amsthm}

\pgfplotsset{compat=1.16}

\begin{document}
%\title{HRIS-Enabled Joint Communications and Multi-Target Tracking with Wideband Transmissions}
\title{Tracking-Aided Multi-User MIMO Communications\\with Hybrid Reconfigurable Intelligent Surfaces}
\author{\IEEEauthorblockN{
Ioannis Gavras and George C. Alexandropoulos 
} 
\\
\IEEEauthorblockA{Department of Informatics and Telecommunications, National and Kapodistrian University of Athens\\Panepistimiopolis Ilissia, 16122 Athens, Greece}
\\
\IEEEauthorblockA{emails: \{giannisgav, alexandg\}@di.uoa.gr}
\vspace{-0.5cm}
}

\maketitle
\begin{abstract}
Hybrid Reconfigurable Intelligent Surfaces (HRISs) constitute a new paradigm that redefines smart metasurfaces, not only offering tunable reflections of incoming signals, but also incorporating signal reception and processing capabilities. In this paper, leveraging the simultaneous dual-functionality of HRISs, we propose a novel framework for tracking-aided multi-user Multiple-Input Multiple-Output (MIMO) communications. In particular, a joint design of the transmit multi-user precoding matrix together with the HRIS reflection and analog combining configurations is presented, with the objective to maximize the accuracy of position estimation of multiple mobile users while  meeting their individual quality-of-service constraints for sensing-aided communications. The Cram\'{e}r-Rao bound for the users' positioning parameters is derived together with a prediction approach based on the extended Kalman filter. Our simulation results showcase the efficacy of the proposed Integrated Sensing And Communications (ISAC) framework over various system configuration parameters.
\end{abstract}
 

\vspace{-0.3cm}
\pagenumbering{gobble}
\section{Introduction}
The forthcoming sixth Generation (6G) of wireless networks is expected to incorporate advanced radar-like sensing capabilities, seamlessly aligning with the late research and standardization trends~\cite{6G-DISAC_mag}. This emerging paradigm, known as Integrated Sensing And Communications (ISAC)~\cite{mishra2019toward}, has gained significant traction, with the latest emphasis being given to definition of use cases, channel modeling, and key enabling technologies\cite{RIS_ISAC,18}. In this regard, the synergy between eXtremely Large (XL) Multiple-Input Multiple-Output (MIMO) systems and frequencies above $6$~GHz presents a promising direction, offering high-resolution angular and range sensing. Additionally, the concept of smart radio environments~\cite{RIS_challenges}, enabled mainly by Reconfigurable Intelligent Surfaces (RISs)~\cite{BAL24}, has been recently presented that allows for programmable control of signal propagation. This feature facilitates various 6G applications, including ISAC~\cite{RIS_ISAC} and simultaneous localization and radio mapping~\cite{kim2023ris}.

Hybrid RISs (HRISs), featuring meta-atoms capable of simultaneously realizing tunable reflection and absorption through embedded power splitters as well as small numbers of Reception (RX) Radio Frequency (RF) chains (as compared to the number of meta-atoms), were recently introduced in~\cite{hybrid_meta-atom} for dual-functional applications and self-configuration~\cite{alexandropoulos2023hybrid}. For instance, the absorbed portion of incoming signals has been utilized for tasks such as channel estimation~\cite{alexandropoulos2020hardware,HRIS_CE} and localization~\cite{RIS_loc,fascista2022ris}, offering a balance between hardware cost, power efficiency, and computational complexity. HRISs are also emerging as a enabling technology for ISAC, leveraging their large aperture size relative to the signal wavelength to enable large-scale sensing capabilities~\cite{alexandropoulos2023hybrid,gavras2024simultaneous}. However, despite their promising features, the HRIS potential in ISAC remains still largely unexplored, especially regarding their role to implement sensing-aided mobile communications.
%especially regarding their role in integrating tracking functionalities. This represents a highly promising application, as HRISs' unique self-configurability, sensing and reflection capabilities, could facilitate tracking-aided wireless communications.

%\vspace{-0.049cm}
In this paper, we present a novel communication-centric bistatic system comprising a MIMO Base Station (BS) and an HRIS that capitalizes on the wideband sensing capability of the latter to enable tracking of the trajectories of mobile User Equipments (UEs), which is then leveraged to optimize the wideband multi-UE DownLink (DL) BS precoding and HRIS reflection configuration under Quality-of-Service (QoS) constraints. In particular, we introduce a state evolution model for the moving UEs and propose a prediction approach based on the Extended Kalman Filter (EKF). In addition, the Cram\'{e}r-Rao Bound (CRB) for the UEs' positioning parameters is derived together with their corresponding Position Error Bound (PEB). Our numerical evaluations highlight the effectiveness of the proposed ISAC framework, demonstrating its tracking-aided communication improvement capability and the inherent trade-offs between communication, localization, and tracking performance.
%novel communication-centric bistatic system comprising a MIMO Base Station (BS) and an HRIS. Considering wideband communication, the objective of the proposed system is to ensure optimal communication performance for the data-requesting User Equipments (UEs) by tracking their trajectories, as they navigate the environment. We derive the Cram\'{e}r-Rao Bound (CRB) for estimating the UEs' positioning parameters and compute the corresponding Position Error Bound (PEB), which serves as the objective of the proposed optimization, subject to a set of communication-oriented Quality of Service (QoS) constraints. 

\textit{Notations:}
Vectors and matrices are represented by boldface lowercase and uppercase letters, respectively. The transpose, Hermitian transpose, and inverse of $\mathbf{A}$ are denoted as $\mathbf{A}^{\rm T}$, $\mathbf{A}^{\rm H}$, and $\mathbf{A}^{-1}$, respectively. $\mathbf{I}_{n}$, $\mathbf{0}_{n}$, and $\boldsymbol{1}_n$ ($n\geq2$) are the $n\times n$ identity and zeros' matrices and the ones' column vector, respectively. $[\mathbf{A}]_{i,j}$ is $\mathbf{A}$'s $(i,j)$-th element, $\|\mathbf{A}\|$ gives its Euclidean norm, and $\text{Tr}\{\mathbf{A}\}$ its trace. $|a|$ ($\Re\{a\}$) returns the amplitude (real part) of complex scalar $a$. $\mathbb{C}$ is the complex number set, $\jmath\triangleq\sqrt{-1}$ is the imaginary unit, $\mathbb{E}\{\cdot\}$ is the expectation operator, and $\mathbf{x}\sim\mathcal{CN}(\mathbf{a},\mathbf{A})$ indicates a complex Gaussian %random 
vector with mean $\mathbf{a}$ and covariance matrix $\mathbf{A}$. %$\exp(\cdot)$ is the exponential function and $\angle\A$ represents the phase information of $\A$. 

\section{System and Channel Models}\label{Sec: System_Sec}
\begin{figure}[!t]
	\begin{center}
    \includegraphics[scale=0.7]{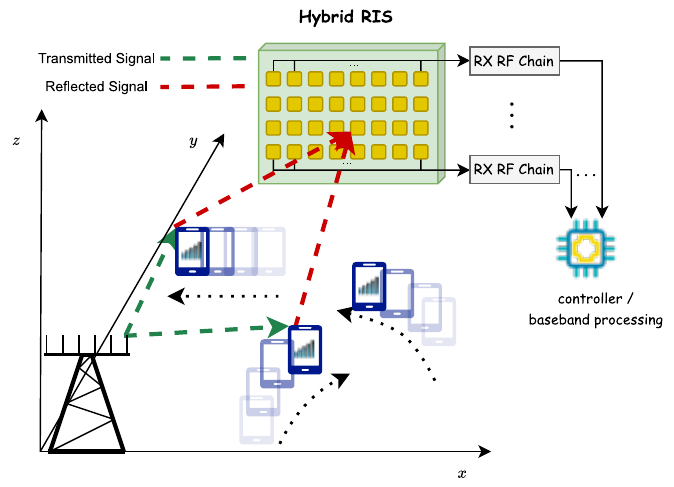}
	\caption{\small{The considered tracking-aided multi-UE wireless communication system deploying an HRIS designed to track the trajectories of the UEs.}}
	\label{fig: system}
	\end{center}
\end{figure}

We consider the wireless communication system of Fig.~\ref{fig: system} consisting of a multi-antenna BS and an HRIS, which wishes to realize efficient wideband communications in the DL direction with $U$ mobile single-antenna UEs, leveraging the sensing capability of the HRIS which is herein designed to track their trajectories. The BS is equipped with an $N_{\rm T}$-element Uniform Linear Array (ULA) and supports fully digital TX BeamForming (BF), whereas the HRIS is modeled as a Uniform Planar Array (UPA) with $N_{\rm H} \triangleq N_{\rm RF}N_{\rm E}$ hybrid meta-atoms, where each column of elements is connected to a distinct RX RF chain ($N_{\rm RF}$ in total, with $N_{\rm H}/N_{\rm RF} \in \mathbb{Z}^+_*$), enabling partial signal absorption (determined by the parameter $\varrho\in[0,1]$ to be detailed in the sequel) for baseband processing\cite{alexandropoulos2023hybrid}. For simplicity, we assume $\lambda/2$ spacing between adjacent elements in both the BS and HRIS, where $\lambda = c/f_c$ is the signal wavelength with $c$ denoting the speed of light and $f_c$ is the operating frequency. We also assume that the BS and HRIS know each other's static positions\cite{RIS_challenges}, and that the controller of the metasurface shares its signal observations and phase configurations with a central processing unit hosted at the BS, which is tasked to implement the ISAC design detailed later on in Section~III.

The OFDM scheme is employed for wideband DL communications, according to which the BS utilizes $K$ subcarriers to transmit $T$ information-bearing OFDM symbols to the $U$ users on a frame-by-frame basis. To ensure the feasibility of BF design for tracking-aided multi-UE communications, we assume $U\leq N_{\rm T}\leq N_{\rm H}$. The symbol matrix per transmission frame is given by $\S\triangleq[\S_1,\ldots,\S_K]\in\Compl^{KU\times T}$, where $\S_k\triangleq[\s_{k,0},\ldots,\s_{k,U}]\in\Compl^{U\times T}$ represents the transmit symbol matrix for all UEs on each $k$th subcarrier ($k=1,\ldots,K$). Notation $\s_{k,u}\in\Compl^{T\times 1}$ indicates the symbol data vector for the $u$th UE ($u=1,\ldots,U$) at the $k$th subcarrier. The compact precoding matrix for all UEs across all subcarriers is defined as $\F_{\rm TX}\triangleq{\rm blkdiag}(\F_1,\ldots,\F_K)\in\Compl^{KN_{\rm T}\times KU}$, where $\F_k\triangleq[\f_{k,0},\ldots,\f_{k,U}]\in\Compl^{N_{\rm T}\times U}$ includes the precoding matrices for all UEs on the $k$th subcarrier with each $\f_{k,u}\in\Compl^{N_{\rm T}\times 1}$ denoting the precoding vector for each $u$th UE. The total transmit power is constrained such that $\mathbb{E}\{|\F_{\rm TX}\S\|^2\} \leq P_{\rm max}$, where $P_{\rm max}$ is the maximum transmission power. 

The simultaneous reflection and sensing functionality of the HRIS is controlled through $N_{\rm H}$ identical power splitters \cite{alexandropoulos2023hybrid,alexandropoulos2020hardware} (their power splitting ratio is modeled by $\varrho$), which divide the power of the impinging signal at each hybrid meta-atom for the respective operations. For the sensing operation, to feed the absorbed portion of the impinging signal to the $N_{\rm RF}$ RX RF chains, the HRIS applies the analog combining matrix $\W_{\rm H}\in\Compl^{N_{\rm H}\times N_{\rm RF}}$, which is modeled as~\cite{gavras2025near}: $[\W_{\rm H}]_{(l-1)N_{\rm E}+n,j}=w_{l,n}$ for $l=j$ ($l,j=1,\ldots,N_{\rm RF}$) and $[\W_{\rm H}]_{(l-1)M_{\rm E}+n,j}=0$ for $l\neq j$, 
where $|w_{l,n}|=1$ ($n=1,\ldots,N_{\rm H}$) for each non-zero element in $\W_{\rm H}$. In addition, the effective HRIS reflection coefficients are represented by $\boldsymbol{\varphi}\in\Compl^{N_{\rm H}\times 1}$ (i.e., the $1-\varrho$ portion of the impiging signal), assuming that $|[\boldsymbol{\varphi}]_n|=1$ $\forall n$.

%The BS applies digital precoding through the $k$-th subcarrier channel-dependent beamforming matrix $\F_k \in \mathbb{C}^{N_{\rm T} \times U}$ to the users' complex-valued symbols $\S_k = [\s_k(1),\ldots, \s_k(T)] \in \mathbb{C}^{U \times T}$. For transmit beamforming to be possible we assume that $U\leq N_{\rm T}$. The total transmit power is constrained such that $\mathbb{E}\{\sum_{k=1}^{K}|\F_k\s(t)\|^2\} \leq P_{\rm max}$, where $P_{\rm max}$ denotes the maximum transmission power.

\subsection{Channel Models}\label{subsec:channel_model}
The $N_{\rm T}\times 1$ complex-valued DL channel for each $k$th subcarrier between the BS and each $u$th mobile UE is modeled in the far field as follows:
\begin{align}\label{eq: DL}
    \h_{{\rm DL},k,u} \triangleq a_u e^{-\jmath2\pi (f_k-f_{{\rm D},u})\tau_u}\a_{\rm BS}(\theta_u),
\end{align}
where $f_k\triangleq f_c+(k-\frac{K+1}{2})\Delta f$ is the frequency of the $k$th subcarrier, $\Delta f$ is the subcarrier spacing, $f_{{\rm D},u}$ is the Doppler shift, $a_u$ is the complex path gain, $\theta_u$ is the azimuth Angle of Departure (AoD) with $\tau_u$ is being the corresponding time delay, and $\a_{\rm BS}(\cdot)\in\Compl^{N_{\rm T}\times 1}$ is the BS steering vector. 
Similarly to \eqref{eq: DL}, the $N_{\rm H}\times N_{\rm T}$ complex-valued Line-of-Sight (LoS)  channel between the BS and the HRIS for each $k$th subcarrier is modeled as follows:
\begin{align}\label{eq: DL_1}
    \H_{{\rm BH},k} \triangleq a_{\rm BR}e^{-\jmath2\pi f_k\tau_{\rm BR}}\a_{\rm H}(\psi_{\rm BR},\phi_{\rm BR})\a_{\rm BS}^{\rm H}(\theta_{\rm BR}),
\end{align}
where $a_{\rm BR}$, $\tau_{\rm BR}$, $\theta_{\rm BR}$, $\psi_{\rm BR}$, and $\phi_{\rm BR}$ denote respectively the complex LoS link attenuation, time delay, azimuth AoD, as well as the azimuth and elevation Angles of Arrival (AoAs) between the BS and HRIS. In addition, $\a_{\rm H}(\cdot)\in\Compl^{N_{\rm H}\times 1}$ is the HRIS steering vector. 

The $N_{\rm H}\times N_{\rm T}$ complex-valued channel, for each $k$th subcarrier, which accounts for single-bounce reflections of the BS's signals from the UEs that are received at the HRIS side, is expressed in the far-field regime as follows:
\begin{align}\label{eq: bistatic}
    \H_{{\rm H},k}\triangleq \sum_{u=1}^Ua_{{\rm H},u}e^{-\jmath2\pi(f_k-f_{{\rm D},u})\tau_{{\rm H},u}}\a_{\rm H}(\psi_u,\phi_u)\a_{\rm BS}^{\rm H}(\theta_u),
\end{align}
where $\a_{{\rm H},u}$ represents the $u$th UE's end-to-end complex path attenuation and $\tau_{{\rm H},u}$ denotes the corresponding time delay. Finally, the parameters $\psi_u$ and $\phi_u$ correspond to the elevation and azimuth AoAs, respectively.

As shown in the Cartesian coordinate system in Fig.~\ref{fig: system}, the BS is positioned at the origin and the HRIS is positioned opposite of the it, with its orientation facing the BS and its first element placed at the  point $\p_{\rm H}\triangleq[x_{\rm H},y_{\rm H},z_{\rm H}]^{\rm T}\in\mathbb{R}^{3\times1}$. Due to the far-field propagation assumption, each $u$th UE is modeled as a point source positioned between the BS and HRIS at the point $\p_{u}\triangleq[x_u,y_u,z_u]^{\rm T}\in\mathbb{R}^{3\times 1}$ $\forall u$. In addition, the BS and HRIS steering vectors with respect to the $u$th UE are given respectively by $\a_{\rm BS}(\theta_u)\triangleq\frac{1}{\sqrt{N_{\rm T}}}[1,e^{\jmath2\pi\sin{\theta_u}},\ldots,e^{\jmath(N_{\rm T}-1)\pi\sin{\theta_u}}]^{\rm T}$ and $\a_{\rm H}(\psi_u,\phi_u)\triangleq\a_{\rm H}^{\rm rows}(\phi_u)\otimes\a_{\rm H}^{\rm cols}(\psi_u)$, where $\a_{\rm H}^{\rm rows}(\cdot)\in\Compl^{N_{\rm RF}\times 1}$ and $\a_{\rm H}^{\rm cols}(\cdot)\in\Compl^{N_{\rm E}\times 1}$ are respectively the steering vectors with respect to the azimuth and elevation AoAs, modeled analogously to $\a_{\rm BS}(\cdot)$. According to this system geometry, the involved AoDs and AoAs in~\eqref{eq: DL}--\eqref{eq: bistatic} can be expressed as:
\begin{align}
    &\nonumber\theta_u = \tan^{-1}\! \left( \frac{y_u}{x_u} \right),\,\phi_u = \tan^{-1}\! \left( \frac{y_u - y_{\rm H}}{x_u - x_{\rm H}} \right),\\
    &\nonumber\psi_u = \tan^{-1}\! \left( \frac{z_u - z_{\rm H}}{\sqrt{(x_{\rm H} - x_u)^2 + (y_{\rm H} - y_u)^2}} \right)\,\forall u,
\end{align}
\begin{align}
\nonumber\theta_{\rm BR}=\psi_{\rm BR}=\tan^{-1}\! \left( \frac{y_{\rm H}}{x_{\rm H}} \right)\!,\,\phi_{\rm BR}=\tan^{-1}\!\left( \frac{z_{\rm H}}{\sqrt{x_{\rm H}^2 + y_{\rm H}^2}} \right).
\end{align}
%\begin{align}
%    &\nonumber\theta_u = \tan^{-1} \left( \frac{y_u}{x_u} \right),\,\phi_u = \tan^{-1} \left( \frac{y_u - y_{\rm H}}{x_u - x_{\rm H}} \right),\\
%    &\nonumber\psi_u = \tan^{-1} \left( \frac{z_u - z_{\rm H}}{\sqrt{(x_{\rm H} - x_u)^2 + (y_{\rm H} - y_u)^2}} \right)\,\forall u,\\
%    &\nonumber\theta_{\rm BR}=\psi_{\rm BR}=\tan^{-1} \left( \frac{y_{\rm H}}{x_{\rm H}} \right),\,\phi_{\rm BR}=\tan^{-1}\left( \frac{z_{\rm H}}{\sqrt{x_{\rm H}^2 + y_{\rm H}^2}} \right).
%\end{align}
To this end, the corresponding time delays are given by:
\begin{align}
    &\nonumber\tau_{{\rm H},u}=(\|\p_{u}\|+\|\p_{\rm H}-\p_{u}\|)/c,\\
    &\nonumber\tau_u =  \|\p_{u}\|/c+\Delta t_u,\, \tau_{\rm BR} =  \|\p_{\rm H}\|_2/c,
\end{align}
where $\Delta t_u$ is the clock bias between the BS and the $u$th UE. Recall that the central processing unit at the BS is connected with the HRIS controller via a reliable link, hence, the clock biases in $\tau_{{\rm H},u}$ $\forall u$ can be compensated. Finally, the complex path gains in~\eqref{eq: DL}--\eqref{eq: bistatic} can be modeled as follows:
\begin{align}
&\nonumber a_u=\frac{\lambda}{4\pi\|\p_{u}\|^2},\, a_{\rm BR}=\frac{\lambda}{4\pi\|\p_{{\rm H}}\|^2},\\ 
&\nonumber a_{{\rm H},u}=\frac{e^{\jmath\omega_u}\lambda^2}{4\pi^{2}\|\p_{u}\|^2\|\p_{\rm H}-\p_{u}\|^2},
\end{align}
where $e^{\jmath\omega_u}$ denotes the $u$th UE's complex-valued reflection coefficient, with $\omega_u$ being uniformly distributed in $[0,2\pi)$ $\forall u$. 

By denoting the velocity of each $u$th UE as $\v_u\triangleq[\dot{x}_u,\dot{y}_u,\dot{z}_u]^{\rm T}\in\mathbb{R}^{3\times 1}$, the corresponding Doppler shift can be computed as $f_{{\rm D},u}=\frac{f_c}{c}(\u_{\rm BS}\v_u+\u_{\rm H}\v_u)$, where $\u_{\rm BS},\u_{\rm H}\in\mathbb{R}^{1\times 3}$ represent the unit vectors along the paths from the BS to the $u$th UE and from that UE to the HRIS. In the case of \eqref{eq: DL}, the Doppler shift is simplified to $\u_{\rm BS}\v_u$.

\subsection{Received Signal Models}
The baseband received signal at each $u$th UE at each $k$th subcarrier, considering $T$ OFDM symbol transmissions from the BS, can be mathematically expressed as follows:
\begin{equation*}%\label{eq:UE_received_signal}
    y_{k,u}\!\triangleq\!\left(\h_{{\rm DL},k,u}\!+\!(1\!-\!\varrho)\h_{{\rm HU},k,u}{\rm diag}(\boldsymbol{\varphi})\H_{{\rm BR},k}\right)\f_{k,u}\s_{k,u}\!+\!n,
\end{equation*}
where $\h_{{\rm HU},k,u}$ represents the $1\times N_{\rm H}$ complex-valued channel vector between the HRIS and each $u$th UE, that is modeled similar to~\eqref{eq: DL}, and $n\sim\mathcal{CN}(0,\sigma^2)$ models the Additive White Gaussian Noise (AWGN). Recall that parameter $\varrho\in[0,1]$ is the common power splitting ratio at the HRIS meta-atoms.

%$t\in\{1,2,\ldots,T\}$
Making the reasonable assumption that the static BS-HRIS channel in~\eqref{eq: DL_1} can be accurately estimated and, thus, completely canceled out, the baseband received signal for each $k$th subcarrier at the outputs of the $N_{\rm RF}$ RX RF chains of the HRIS can be mathematically formulated as follows: 
\begin{align}\label{eq: y_ref}
    \Y_k = [\y_k(1),\ldots,\y_k(T)] \triangleq \varrho\W_{\rm H}^{\rm H}\H_{{\rm H},k}\F_k\S_k+\N,
\end{align}
where $\N \triangleq [\n(1),\ldots,\n(T)]\in\Compl^{N_{\rm RF}\times T}$, with $\n(t)\sim\mathcal{CN}(0,\sigma^2\I_{N_{\rm RF}})$ $\forall t=1,\dots,T$, denotes the matrix with the AWGN vectors per received OFDM symbol.

\section{Tracking-Aided Multi-UE Communications}
In this section, we present the evolution model for the mobile UEs and introduce an EKF-based approach to track their positions and velocities in space. We start by deriving the CRB for the UEs' positioning vector, and then propose a joint communication and tracking optimization framework.

\subsection{PEB Analysis}
We define the channel- and location-domain parameter vectors within a frame as $\boldsymbol{\eta}\triangleq[\x,\y,\z,\dot{\x},\dot{\y},\dot{\z}]^{\rm T}\in\mathbb{R}^{6U\times 1}
$ and $\widetilde{\boldsymbol{\eta}}\triangleq[\boldsymbol{\theta},\boldsymbol{\psi},\boldsymbol{\phi},\boldsymbol{\tau}_{\rm H}]\in\mathbb{R}^{4U\times 1}$, respectively, where $\x\triangleq[x_1,\ldots,x_U]$, $\y\triangleq[y_1,\ldots,y_U]$, $\z\triangleq[z_1,\ldots,z_U]$, $\boldsymbol{\theta}\triangleq[\theta_1,\ldots,\theta_U]$, $\boldsymbol{\psi}\triangleq[\psi_1,\ldots,\psi_U]$, $\boldsymbol{\phi}\triangleq[\phi_1,\ldots,\phi_U]$, $\boldsymbol{\tau}_{\rm H}\triangleq[\tau_{{\rm H},1},\ldots,\tau_{{\rm H},u}]$, $\dot{\x}\triangleq[\dot{x}_1,\ldots,\dot{x}_U]$, $\dot{\y}\triangleq[\dot{y}_1,\ldots,\dot{y}_U]$, and $\dot{\z}\triangleq[\dot{z}_1,\ldots,\dot{z}_U]$. It is evident from \eqref{eq: y_ref}'s inspection that, for a coherent channel block involving $T$ unit-powered symbol transmissions $\forall t=1,2,\ldots,T$, yields $T^{-1}\S\S^{\rm H}=\I_{KU}$, therefore, the received signal at the outputs of the RX RF chains of the HRIS node can be modeled as $\y_k\triangleq{\rm vec}\{\Y_k\}\sim\mathcal{CN}(\boldsymbol{\mu}_k,\sigma^2\I_{N_{\rm RF}T})$ with mean $\boldsymbol{\mu}_k \triangleq {\rm vec}\{\varrho\W_{\rm H}^{\rm H}\H_{{\rm H},k}\F_k\S_k\}$. The $\J\in\mathbb{R}^{6U\times6U}$ Fisher Information Matrix (FIM) of $\boldsymbol{\eta}$ can be computed analogously to\cite{kay1993fundamentals} as follows:
\begin{align}\label{eq: FIM}
    [\J]_{i,j}=\frac{2T\varrho^2}{\sigma^2}\sum_{k=1}^K\Re\left\{{\rm Tr}\left\{\frac{\partial\Bar{\boldsymbol{\mu}}_k^{\rm H}}{\partial\eta_i}\frac{\partial\Bar{\boldsymbol{\mu}}_k}{\partial\eta_j}\right\}\right\},
\end{align}
where $i,j=1,\ldots,6U$ and $\Bar{\boldsymbol{\mu}}_k\triangleq\W_{\rm H}^{\rm H}\H_{{\rm H},k}\F_k$. Assuming DL transmissions in $M$ consecutive frames, where each frame contains $T$ OFDM symbols transmitted by the BS, and focusing on the positioning and tracking performance up to the $m$th frame ($m=1,\ldots,M$), we derive the FIM $\widetilde{\J}_m\in\mathbb{R}^{4U\times4U}$ as follows \cite[Chap. 3.8]{kay1993fundamentals}:
\begin{align}\label{eq: TPEB}
    \widetilde{\J}_m = \T^{\rm T}\J_m\T+\widetilde{\J}_{m-1},
\end{align}
where $\T^{\rm T}\J_m\T$ and $\widetilde{\J}_{m-1}$ represent the FIMs with respect to $\widetilde{\boldsymbol{\eta}}_m$ and the prior knowledge, respectively, and the
transformation matrix $\T\in\mathbb{R}^{6U\times4U}$ can be expressed as a Jacobian with $[\T]_{i,j}=\partial[\widetilde{\boldsymbol{\eta}}]_i/\partial[{\boldsymbol{\eta}}]_j$. Note that $\boldsymbol{\eta}_m$ and $\widetilde{\boldsymbol{\eta}}_m$ are the channel- and location-domain parameter vectors for the $m$th frame and $\J_m$ is computed analogously to \eqref{eq: FIM} $\forall m$. At the first frame, $\widetilde{\J}_0$ is equal to a zero matrix, since no prior knowledge is available yet. Combining all the above to quantify the position accuracy up to each $m$th frame, we adopt the PEB as our estimation performance metric, which can be computed as follows:
\begin{align}\label{eq: PEB}
    {\rm PEB}(\F_{\rm TX},\W_{\rm H};\widetilde{\boldsymbol{\eta}}_m)=\sqrt{{\rm Tr}\left\{\widetilde{\J}^{-1}_m\right\}}.
\end{align}

\subsection{State Evolution Model}
We assume that all UEs move with constant velocity from frame to frame. Under the considered Cartesian coordinate system, each $u$th UE has an initial location $(x_u^0,y_u^0,z_u^0)$ and velocity $(\dot{x}_u^0,\dot{y}_u^0,\dot{z}_u^0)$, and we define each $u$th state vector for each $m$th frame as $\xi_u^m\triangleq[x_u^m,y_u^m,z_u^m,\dot{x}_u^m,\dot{y}_u^m,\dot{z}_u^m]$. To track the locations of the mobile UEs, we introduce a state prediction model and a measurement model similar to~\cite{wang2024resource}. These models, together with the previous CRB analysis, allow for the assessment of the tracking performance.

\subsubsection{System Dynamics} The motion evolution of each $u$th
UE between adjacent frames $m$ and $m-1$ is given by:
\begin{align}
    \xi_u^m=\F_{\xi}\xi_u^{m-1}+\boldsymbol{\rho}_{u}^{m-1}
\end{align}
where $\F_{\xi}$ is the transition matrix between two frames, which can be set as follows~\cite{wang2024resource}:
\begin{align}
    \F_{\xi}=\I_3\otimes\begin{bmatrix}
        1 & T_s \\ 0 & 1
    \end{bmatrix}
\end{align}
with $T_s$ being the time interval between the adjacent frames and $\boldsymbol{\rho}_{u}^{m-1}$ models the process noise at the $(m-1)$th frame, which follows a Gaussian distribution with zero mean and covariance $\P_u$ given by:
\begin{align}
    \P_u = \dot{\sigma}\I_3\otimes\begin{bmatrix}
        \frac{1}{3}T_s^3 & \frac{1}{2}T_s^2 \\ \frac{1}{2}T_s^2 & T_s.
    \end{bmatrix}
\end{align}
In this expression, $\dot{\sigma}$ is the process noise level of the UEs.

\subsubsection{Measurement Model} The positions of the UEs can be tracked using their AoDs, AoAs, and time delays at the HRIS end. Specifically, the evolution of each $u$th UE's parameters at each $m$th frame can be modeled as~\cite{wang2024resource}:
\begin{align}
    \z_u^m = \g(\xi_u^m)+\Bar{\boldsymbol{\rho}}_u^m,
\end{align}
where $\g(\xi_u^m)\triangleq[\theta_u^m,\psi_u^m,\phi_u^m,\tau_{{\rm H},u}^m]^{\rm T}\in\mathbb{R}^{4}$ indicates the mapping function, which is given via Section~\ref{subsec:channel_model} as follows:
\begin{align}
    &\nonumber\theta_u^m = \tan^{-1} \left( \frac{y_u^m}{x_u^m} \right),\,\psi_u^m = \tan^{-1} \left( \frac{z_u^m - z_{\rm H}}{r_u^m} \right)\\
    &\nonumber\phi_u^m = \tan^{-1} \left( \frac{y_u^m - y_{\rm H}}{x_u^m - x_{\rm H}} \right),\,\tau_{{\rm H},u}^m =  \frac{\|\p_{u}^m\|+\|\p_{\rm H}-\p_{u}^m\|}{c},
\end{align}
where $r_u^m=\sqrt{(x_{\rm H} - x_u^m)^2 + (y_{\rm H} - y_u^m)^2}$. In this expression, $\p_{u}^m\triangleq[x_u^m,y_u^m,z_u^m]$ and $\Bar{\boldsymbol{\rho}}_u^m$ is the measurement error following a Gaussian distribution with zero mean and covariance matrix given by:
\begin{align}
    \boldsymbol{\Sigma}_u^m = {\rm diag}\left(\sigma^2_{\theta_u^m},\sigma^2_{\psi_u^m},\sigma^2_{\phi_u^m},\sigma^2_{\tau_{{\rm H},u}^m}\right)
\end{align}
with $\sigma^2_{\theta_u^m},\sigma^2_{\psi_u^m},\sigma^2_{\phi_u^m}$ and $\sigma^2_{\tau_{{\rm H},u}^m}$ being the CRBs of the AoAs, AoDs, and time delay of each $u$th target at each $m$th frame, which can be directly obtained from~\eqref{eq: TPEB} as follows:
\begin{align}
&\nonumber\sigma^2_{\theta_u^m}=\left[\widetilde{\J}_m^{-1}\right]_{u,u}, \sigma^2_{\psi_u^m}=\left[\widetilde{\J}_m^{-1}\right]_{U+u,U+u},\\
    &\sigma^2_{\phi_u^m}=\left[\widetilde{\J}_m^{-1}\right]_{2U+u,2U+u}, \sigma^2_{\tau_{{\rm H},u}^m}=\left[\widetilde{\J}_m^{-1}\right]_{3U+u,3U+u}.
\end{align}

\subsubsection{EKF-Based Prediction} After receiving the reflected echoes, the AoAs, AoDs, and time delays are estimated. The EKF is then used to predict the states of the UEs, which are also tracked by integrating these predictions within the measurements. Notably, we estimate only the UEs' positions, but, by leveraging the EKF, we can also track their velocity. Finally, the EKF process consists of the following steps~\cite{wang2024resource}:
\begin{itemize}
    \item State Prediction:
    \begin{align}
        \hat{\xi}_u^{m|m-1}=\F_{\xi}\hat{\xi}_u^{m-1}.
    \end{align}
    \item Mean-Squared Error Matrix Prediction:
    \begin{align}
        \A_u^{m|m-1}=\F_{\xi}\A_u^{m-1}\F_{\xi}^{\rm T}+\P_u.
    \end{align}
    \item Kalman Gain Calculation:
    \begin{align}
        \nonumber\K_{u}^m=&\A_u^{m|m-1}\left(\hat{\Q}_u^m\right)^{\rm T}\\&\times\left(\hat{\boldsymbol{\Sigma}}_u^m+\hat{\Q}_u^m\A_u^{m|m-1}\left(\hat{\Q}_u^m\right)^{\rm T}\right)^{-1}.
    \end{align}
    \item State Tracking:
    \begin{align}
        \hat{\xi}_u^{m}=\hat{\xi}_u^{m|m-1}+\K_u^m\left(\z_u^m-\g\left(\hat{\xi}_u^{m|m-1}\right)\right)
    \end{align}
    \item Mean-Squared Error Matrix Update:
    \begin{align}
        \A_u^m=\left(\I_6-\K_u^m\hat{\Q}_u^m\right)\A_u^{m|m-1}.
    \end{align}
\end{itemize}
It is noted that $\hat{\Q}_u^m$ and $\hat{\boldsymbol{\Sigma}}_u^m$ are the Jacobian matrix of $\g(\xi_u^m)$ and the covariance matrix
of the measurement error evaluated at the predictive state $\hat{\xi}_u^m$, respectively.

%\vspace{-0.5cm}
\subsection{Proposed BS and HRIS Design}
Our objective is to jointly optimize the BS precoding matrix a well as the HRIS reflecting and analog combining parameters to meet communication QoS constraints for all mobile UEs in the system, while tracking their positions in time (i.e., over multiple frames) to optimize DL transmissions. %Commonly, ensuring each UE's SINR meets a predefined threshold is critical for communication performance. 
The SINR-based communication objective for each $u$th UE at each $k$th subcarrier per OFDM symbol can be calculated as follows:
\begin{align}
    &{\rm SINR}_{k,u}=\frac{\left|\h_{{\rm dir},k,u}\f_{k,u}\right|^2}{\sum_{i=0,i\neq u}^{U}\left|\h_{{\rm dir},k,u}\f_{k,i}\right|^2+\sigma^2},
\end{align}
where $\h_{{\rm dir},k,u}\triangleq\h_{{\rm DL},k,u}+(1-\rho)\h_{{\rm HU},k,u}{\rm diag}(\boldsymbol{\varphi})\H_{{\rm BR},k}$, which is composed at each $m$th frame $\forall k,u$ using the state estimation $\hat{\xi}_u^m$ for each $u$th UE. From the sensing perspective, we focus on minimizing the PEB in~\eqref{eq: PEB} at each $m$th frame. However, to design $\F_{\rm TX}$ and $\W_{\rm H}$ directly optimizing this expression is cumbersome, since high-dimensional constraints with numerous linear matrix inequalities need to be incorporated. To deal with this complexity, we deploy \cite{gavras2025near}'s approach and minimize instead the lower bound ${\rm Tr}\left\{\widetilde{\J}^{-1}_m\right\} \geq 1/{\rm Tr}\left\{\widetilde{\J}_m\right\}$; we do so at each $m$th frame. Overall, our design objective for each $m$th frame is mathematically expressed as follows:
\begin{align}
        \mathcal{OP}:\nonumber&\underset{\substack{\F_{\rm TX},\W_{\rm H},\boldsymbol{\varphi}}}{\max} \,\,{\rm Tr}\left\{\widetilde{\J}_m\right\}\\
        &\nonumber\text{\text{s}.\text{t}.}\,\sum_{k=1}^K{\rm SINR}_{k,u}\geq\gamma_{u}\,\forall u,\,\left\|\F_{\rm TX}\right\|^2 \leq P_{\rm{\max}},
        \\&\nonumber\,\quad\,|w_{l,n}|=1\,\,\text{and}\,\,|[\boldsymbol{\phi}]_n|=1\,\forall l,n.
\end{align}
Here, $\gamma_{u}$ represents the SINR threshold for the $u$th user. Since the constraints render $\mathcal{OP}$ highly non-convex, solving it directly is challenging. To treat this, we adopt an alternating optimization approach. Specifically, for a given $\W_{\rm H}$ and $\boldsymbol{\varphi}$, $\mathcal{OP}$ can be relaxed to the following problem:
\begin{align}
        &\mathcal{OP}_1:\nonumber\underset{\substack{\{\widetilde{\F}_{k,u}\}_{\forall k,u}}}{\max} \,\,\sum_{i=1}^{4U}\sum_{k=1}^K\sum_{u=1}^U\Re\left\{{\rm Tr}\left\{\B_{k,i}\widetilde{\F}_{k,u}\right\}\right\}\\
        &\nonumber\text{\text{s}.\text{t}.}\,\sum_{k=1}^K{\rm Tr}\left(\C_{k,u}\widetilde{\F}_{k,u}\right)-\gamma_{u}\sum_{i=1,i\neq u}^U{\rm Tr}\left(\C_{k,u}\widetilde{\F}_{k,i}\right)\geq\gamma_{u}\sigma^2,
        \\&\nonumber\,\quad\, \sum_{k=1}^K\sum_{u=1}^U\|\widetilde{\F}_{k,u}\|^2\leq P_{\rm max},\,\widetilde{\F}_{k,u}\succeq0\,\forall k,u,
\end{align}
where $\B_{k,i}\triangleq\frac{\partial\H_{{\rm H},k}^{\rm H}}{\partial\widetilde{\eta}_i}\W_{\rm H}\W_{\rm H}^{\rm H}\frac{\partial\H_{{\rm H},k}}{\partial\widetilde{\eta}_i}$ and $\C_{k,u}\triangleq\h_{{\rm dir},k,u}^{\rm H}\h_{{\rm dir},k,u}$. Note that we have relaxed $\mathcal{OP}$ via the incorporation of $\F_k\F_k^{\rm H}=\sum_{u=0}^{U}\f_{k,u}\f_{k,u}^{\rm H}=\sum_{u=1}^U\widetilde{\F}_{k,u}$, replacing the non-convex rank constraints with positive SemiDefinite (SD) ones, i.e., a series of SD relaxations. To this end, $\mathcal{OP}_1$ is in a convex form, thus, it can be efficiently solved with common convex solvers (e.g., CVX). After performing this optimization, each precoder $\f_{k,u}$ is designed by simply using the principal singular value of $\widetilde{\F}_{k,u}$. As demonstrated in~\cite{alexandropoulos2025extremely}, this approach yields precoders inherently satisfying the corresponding rank-one constraints.

For a given matrix $\F_{\rm TX}$ and vector $\boldsymbol{\phi}$, we leverage $\W_{\rm H}$'s partially-connected structure~\cite{alexandropoulos2023hybrid} to reformulate $\mathcal{OP}$ as:
\begin{align}
        &\mathcal{OP}_2:\nonumber\underset{\substack{\{\W_{l}\}_{l=1}^{N_{\rm RF}}}}{\max} \,\,\sum_{i=1}^{4U}\sum_{k=1}^K\sum_{l=1}^{N_{\rm RF}}\Re\left\{{\rm Tr}\left\{[\D_{k,i}]_{k_l:lN_{\rm E},k_l:lN_{\rm E}}\W_l\right\}\right\}\\
        &\nonumber\text{\text{s}.\text{t}.}\,{\rm diag}(\W_l)=\boldsymbol{1}_{N_{\rm H}},\,\W_l\succeq 0\, \forall l,
\end{align}
where $\D_{k,i}\triangleq\frac{\partial\H_{{\rm H},k}}{\partial\widetilde{\eta}_i}\F_k\F_k^{\rm H}\frac{\partial\H_{{\rm H},k}^{\rm H}}{\partial\widetilde{\eta}_i}$. Note that $\W_{\rm H}\W_{\rm H}^{\rm H}$ forms a block diagonal matrix, simplifying $\mathcal{OP}$, where $\W_l\triangleq\W_{\rm H}(k_l:lN_{\rm E},l)\W_{\rm H}^{\rm H}(k_l:lN_{\rm E},l)$ with $k_l\triangleq(l-1)N_{\rm E}+1$ $\forall l$. Each rank-one constraint can be then replaced with a positive SD one using SDR. $\mathcal{OP}_2$ can be efficiently solved like $\mathcal{OP}1$, with $\W_{\rm H}(k_l:lN_{\rm E},l)$ $\forall l$ constructed from the phase shifts of $\W_l$'s principal singular vector.

The HRIS reflection phase configuration vector $\boldsymbol{\varphi}$ can be finally designed via the following optimization problem that uses the definition $\boldsymbol{\varphi}=e^{\jmath\boldsymbol{\upsilon}}$ with the vector $\boldsymbol{\upsilon}$ including the $N_{\rm H}$ tunable reflection coefficients:
\begin{align}
        \mathcal{OP}_3:&\,\nonumber\underset{\substack{\boldsymbol{\upsilon}}}{\max} \quad\sum_{k=1}^K\sum_{u=1}^U\|\h_{{\rm dir},k,u}(e^{\jmath\boldsymbol{\upsilon }})\f_{k,u}\|^2\\
        &\nonumber\text{\text{s}.\text{t}.}\,\, -\frac{\pi}{2}\leq[\boldsymbol{\upsilon}]_n\leq\frac{\pi}{2}\,\forall n=1,\ldots,N_{\rm H}.
\end{align}
This problem can be efficiently solved with any of the available optimization approaches for conventional RISs~\cite{Tsinghua_RIS_Tutorial}.

\begin{figure*}[!t]
  \begin{subfigure}[t]{0.33\textwidth}
  \centering
    \includegraphics[width=\textwidth]{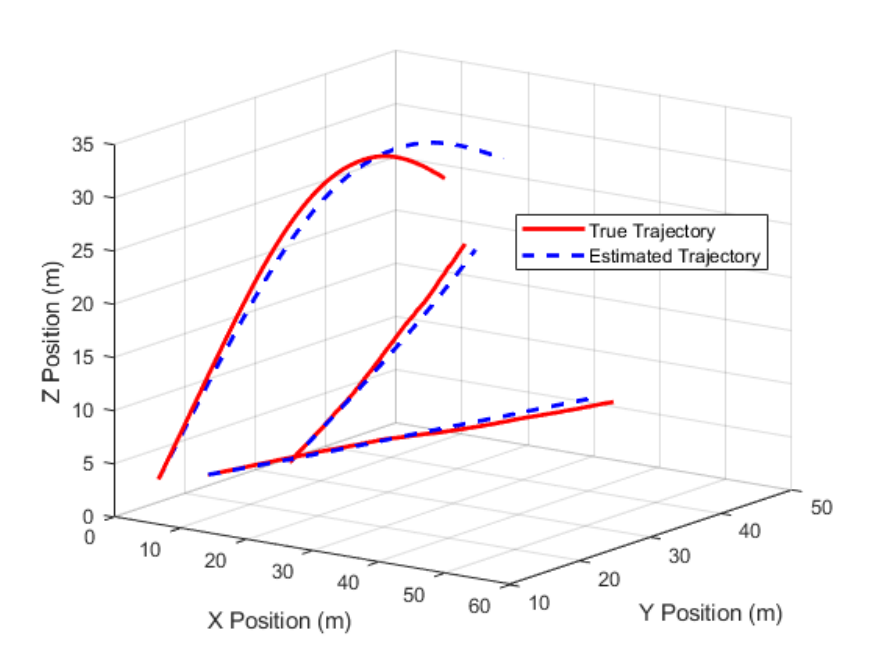}
    \caption{True and Estimated Trajectories.}
    \label{fig:track}
  \end{subfigure}\hfill
  \begin{subfigure}[t]{0.33\textwidth}
  \centering
    \includegraphics[width=\textwidth]{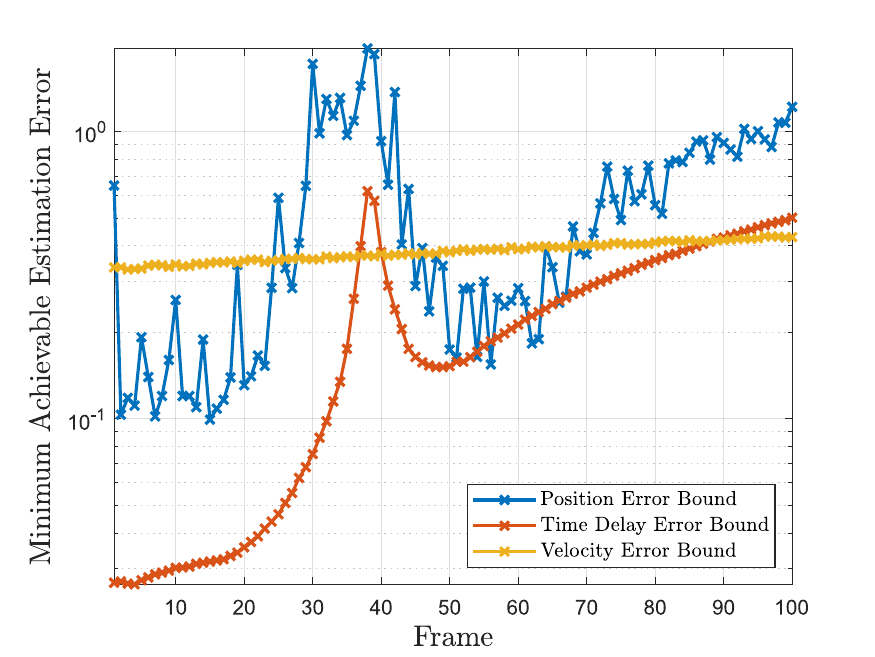}
    \caption{Tracking Performance.}
    \label{fig:Lower_bound}
  \end{subfigure}\hfill
  \begin{subfigure}[t]{0.33\textwidth}
  \centering
    \includegraphics[width=\textwidth]{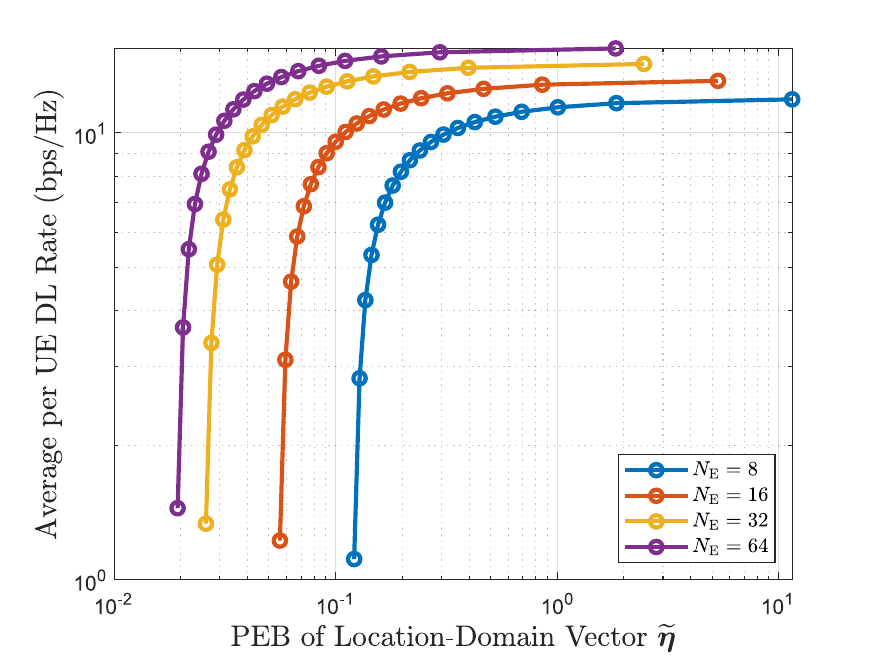}
    \caption{Sensing vs. Communications.}
    \label{fig:tradeoff}
  \end{subfigure}
  \caption{\small{Localization and achievable rate performances focusing on a single UE, considering $\gamma_{u} = 10$ dB $\forall u$ and $P_{\rm max} = 15$~dBm.}}\vspace{-0.4cm}
  \label{fig:Sim}
\end{figure*}

The entire tracking-assisted BS precoding and HRIS configuration approach summarizes as follows: At each $m$th frame, the BS transmits $T$ OFDM symbols and the HRIS captures their reflections from the UEs to estimate the UEs' location vector $\widetilde{\boldsymbol{\eta}}_m$. It then computes the FIM matrix $\widetilde{\J}_m$, performs EKF prediction, and updates the UEs' state vectors $\{\xi_u^m\}_{u=1}^U$. $\mathcal{OP}$ is then iteratively solved by sequentially computing the BS precoder $\F_{\rm TX}$, HRIS combining matrix $\W_{\rm H}$, and the HRIS reflection vector $\boldsymbol{\phi}$ via solving $\mathcal{OP}_1$, $\mathcal{OP}_2$, and $\mathcal{OP}3$, respectively, until convergence or a maximum number of iterations has been reached.

\section{Numerical Results and Discussion}
In this section, we numerically evaluate the HRIS-enabled tracking-aided multi-UE communications framework. We have simulated $M=100$ frames of a wideband setup centered at the $20$~GHz frequency, with $K=32$ subcarriers of $\Delta f=120$ kHz frequency spacing, where each coherent channel block spanned $T=200$ OFDM transmissions. For the estimation of the UEs' localization parameters, we have deployed the approach in~\cite{fascista2022ris}. We have considered $U=3$ mobile UEs initially randomly positioned within the range $[0,50]$ in meters along the $y$-axis, and the range $[0,10]$ in meters along both the $x$- and the $z$-axes; the HRIS was placed at $\p_{\rm RIS}=[0,50,5]$. The velocities of the UEs in each spatial direction were randomly initialized within the range $[1,10]$ in meters per second, and remained constant throughout the entire tracking process. We have considered a BS equipped with ULA of $N_{\rm T}=16$ antennas and an HRIS with $N_{\rm RF} = 5$ RX RF chains each connected to $N_{\rm E} = 8$ hybrid meta-atoms, unless otherwise stated. We have set AWGN's variance as $\sigma^2 = -100$ dBm, and the channel coefficients $\beta_k$ $\forall$$k$ were randomly selected with unit amplitude. The process noise level at the UEs was set to $\dot{\sigma}=0.1$ and the time step between consecutive frames was set as $T_s=0.1$.

The communication and tracking performance of the proposed ISAC approach is illustrated in Fig.~\ref{fig:Sim}, considering the threshold $\gamma_{u} = 10$ dB $\forall u$ and with a maximum transmit power of $P_{\rm max} = 15$~dBm at the BS. Specifically, Fig.~\ref{fig:track} depicts both the estimated and the true trajectories of the UEs, showcasing the efficient tracking of all three UEs while maintaining compliance with the individual QoS communication constraints. The close alignment between the estimated and true trajectories underscores the robustness of the proposed framework, demonstrating the system's capability to effectively integrate current measurements with EKF-based predictions to provide accurate localization within the given data communication horizon. In Fig.~\ref{fig:Lower_bound}, we present the minimum achievable estimation error, which includes the combined position, time delay, and velocity error bounds for all the UEs. These bounds are directly derived from the FIM matrices in~\eqref{eq: FIM} and~\eqref{eq: TPEB}, and correspond to the tracking scenario illustrated in Fig.~\ref{fig:track}. As shown, the PEB is relatively high in the first frame. This is expected, as it represents the initial estimation before any EKF-based prediction is applied. Throughout the tracking process, error bounds generally increase, which is anticipated due to the constant velocity of the UEs in all spatial directions. However, as the UEs move farther from the BS and HRIS, localization becomes more challenging because of the induced path loss. Specifically, in the frame $m=40$ of the tracking sequence, both the position and time delay error bounds reach peaks. This is due to the top UE in Fig.~\ref{fig:track} beginning to turn, which complicates the EKF prediction. However, as the tracking continues, the EKF adjusts and refines its estimates using the available measurements, leading to improved performance. It is also important to note that our approach focuses solely on position estimation of the UEs, with a velocity prediction generated using a constant velocity model. Consequently, the error bound remains relatively constant, with the slight increase being  primarily attributed to the process noise.

In Fig.~\ref{fig:tradeoff}, we illustrate the trade-off between the PEB of the location-domain parameters and the average per user DL rate, considering different numbers of HRIS elements per RX RF chain. The results refer to the last tracking frame of the scenario depicted in Fig.~\ref{fig:track}. The trade-offs have been evaluated for various values of the absorption coefficient $\varrho\in[0,1]$. It can be observed that, when the number of hybrid meta-atoms at the HRIS increases, both the localization and communication performances improve, with the former experiencing the greater gain. It is also shown that, even when HRIS reflects most of the signal, thus, promoting data communications, satisfactory sensing performance is achieved. This behavior confirms the efficacy of our approach in integrating tracking with communications.

\section{Conclusion}
In this paper, a novel wideband HRIS-enabled tracking-aided multi-UE MIMO communication system has been presented. We derived a novel CRB analysis for assessing the system's estimation accuracy, and designed a state evolution model for EKF-based tracking. A joint design of the BS multi-UE precoding matrix together with the HRIS reflection and analog combining configurations was devised, with the objective to minimize the PEB of the UEs while meeting UE QoS constraints for sensing-aided DL communications. The presented numerical results showcased our framework's high-precision tracking capability, demonstrating the trade-off between localization and communications as a function of the HRIS's absorption coefficients.

%\vspace{-0.5cm}
\section*{Acknowledgments}
This work has been supported by the SNS JU projects TERRAMETA and 6G-DISAC under the EU's Horizon Europe research and innovation programme under Grant Agreement numbers 101097101 and 101139130, respectively. TERRAMETA also includes top-up funding by UKRI under the UK government's Horizon Europe funding guarantee.

%\end{NoHyper}
\bibliographystyle{IEEEtran}
\bibliography{ms}

\end{document}